\documentclass[aps,pra,floats,floatsfix,twocolumn,superscriptaddress,showpacs]{revtex4-1}

\usepackage{dcolumn} 
\usepackage{bm} 
\usepackage{amsthm}
\usepackage{latexsym}
\usepackage{float}
\usepackage{amsfonts}
\usepackage{amsmath}
\usepackage{amssymb}
\usepackage{color,graphicx}
\usepackage{tabularx}
\usepackage{balance}
\begin{document}

\title{Soliton repetition rate in a silicon-nitride microresonator}

\author{Chengying Bao}
\email{bao33@purdue.edu}
\affiliation{School of Electrical and Computer Engineering, Purdue University, 465 Northwestern Avenue, West Lafayette, IN 47907-2035, USA}

\author{Yi Xuan}
\affiliation{School of Electrical and Computer Engineering, Purdue University, 465 Northwestern Avenue, West Lafayette, IN 47907-2035, USA}
\affiliation{Birck Nanotechnology Center, Purdue University, 1205 West State Street, West Lafayette, Indiana 47907, USA}

\author{Cong Wang}
\affiliation{School of Electrical and Computer Engineering, Purdue University, 465 Northwestern Avenue, West Lafayette, IN 47907-2035, USA}

\author{Jose A. Jaramillo-Villegas}
\affiliation{School of Electrical and Computer Engineering, Purdue University, 465 Northwestern Avenue, West Lafayette, IN 47907-2035, USA}
\affiliation{Facultad de Ingenier\'{i}as, Universidad Tecnol\'{o}gica de Pereira, Pereira, RI 66003, Colombia}

\author{Daniel E. Leaird}
\affiliation{School of Electrical and Computer Engineering, Purdue University, 465 Northwestern Avenue, West Lafayette, IN 47907-2035, USA}

\author{Minghao Qi}
\affiliation{School of Electrical and Computer Engineering, Purdue University, 465 Northwestern Avenue, West Lafayette, IN 47907-2035, USA}
\affiliation{Birck Nanotechnology Center, Purdue University, 1205 West State Street, West Lafayette, Indiana 47907, USA}

\author{Andrew M. Weiner}
\affiliation{School of Electrical and Computer Engineering, Purdue University, 465 Northwestern Avenue, West Lafayette, IN 47907-2035, USA}
\affiliation{Birck Nanotechnology Center, Purdue University, 1205 West State Street, West Lafayette, Indiana 47907, USA}

\begin{abstract}
The repetition rate of a Kerr comb comprising a single soliton in an anomalous dispersion silicon nitride microcavity is measured as a function of pump frequency tuning.  The contributions from the Raman soliton self-frequency shift (SSFS) and from thermal effects are evaluated both experimentally and theoretically; the SSFS is found to dominate the changes in repetition rate.  The relationship between the changes in repetition rate and pump frequency detuning is found to be independent of the nonlinearity coefficient and dispersion of the cavity.  Modeling of the repetition rate change by using the generalized Lugiato-Lefever equation is discussed; the Kerr shock is found to have only a minor effect on repetition rate for cavity solitons with duration down to $\sim$50 fs.
\end{abstract}

\maketitle

Optical frequency combs (OFCs) consist of a series of discrete, evenly spaced spectral lines, whose frequency is $\nu_{n}$=$nf_r+f_0$, where $f_r$ is the repetition rate and $f_0$ is the carrier-envelope offset frequency \cite{Cundiff_RMP2003colloquium}. OFCs have become an indispensable tool in optical metrology, high precision spectroscopy, optical atomic clock etc. For many applications, highly stabilized OFCs are desired. OFCs based on mode-locked lasers have been shown to exhibit frequency uncertainty at the level of 10$^{-19}$ \cite{Ma_Science2004optical}. Microresonator based Kerr combs \cite{Kippenberg_Science2011microresonator} show potential as a compact replacement for mode-locked laser combs, but they generally do not reach similar stability. Since the integer $n$ is generally large, the frequency of an individual comb line is extremely sensitive to the fluctuation of $f_r$. Hence, investigating the repetition rate of Kerr combs is important to the improvement of Kerr combs. Thermal effects and mechanical stretching have been exploited to stabilize the $f_r$ of Kerr combs \cite{Kippenberg_PRL2008full,Diddams_PRX2013mechanical,Wong_SR2015low,Diddams_NP2016phase}. Recently, cavity solitons (CSs) have been demonstrated in microresonators \cite{Kippenberg_NP2014temporal,Kippenberg_Science2016photonic,Vahala_Optica2015soliton,Weiner_OE2016intracavity,
Gaeta_OL2016Thermal}. Unlike other coherent operation regimes of Kerr combs that are not soliton-like, CSs exhibit a soliton self-frequency shift (SSFS) induced by stimulated Raman scattering (SRS) \cite{Kippenebrg_PRL2016raman,Vahala_Optica2015soliton,Weiner_OE2016intracavity,Vahala_Optica2016Spatial}. The center frequency shift arising through SSFS can affect $f_r$ via dispersion. This constitutes a new mechanism, in addition to thermal effects, which contributes to changes in the repetition rate. Therefore, it is important to compare the relative importance of SSFS and thermal contributions to repetition rate changes (denoted $\Delta f_r$) in the CS regime. In this Letter, we use frequency comb assisted spectroscopy \cite{Kippenberg_NP2009frequency,Weiner_OE2016thermal} to measure the $\Delta f_r$ for a 227 GHz soliton Kerr comb generated from a silicon-nitride (SiN) microresonator. We find that $f_r$ can vary $\sim$25 MHz as the pump frequency is tuned, while maintaining the single soliton state. The contributions from SSFS and thermal effects are isolated and compared, and the SSFS is found to dominate.

The Lugiato-Lefever equation (LLE) is now widely used to model Kerr combs \cite{Coen_OL2013modeling}. However, it has seen little use for modeling of changes in repetition rate and pulse timing. Whether the LLE is capable of capturing such changes is an interesting question, since pulse propagation is averaged in the derivation of LLE. Changes in pulse timing have been modeled and compared to experiment quite recently in the normal dispersion regime \cite{Weiner_arXiv2016second}. Here, we further show LLE is capable of modeling of the changes in $f_r$ we observe in experiments in the CS regime. In further simulations, we show that the Kerr shock has a minor effect on $f_r$ for pulses as short as 50 fs.


The optical path length and the group velocity determine $f_r$ of a pulse train emitted from a cavity. The group velocity in turn depends on the center frequency of the pulses. For mode-locked lasers, the center frequency is constrained by the net gain spectrum; a weak modulation of the pump power can only change the center frequency slightly by several tens GHz \cite{Cundiff_OL2007quantitative,Cundiff_OL2014pulse}. Hence, this center frequency shift will not change $f_r$ significantly; a piezoelectric actuator \cite{Cundiff_RMP2003colloquium} or intracavity electro-optic modulator \cite{Cundiff_OL2005mode} is usually used to change the optical path length to control $f_r$. Unlike mode-locked lasers and non-soliton coherent regimes of Kerr combs, the center frequency of CSs is strongly influenced by the interaction with the waveguide via SRS \cite{Kippenebrg_PRL2016raman}. Tuning the pump frequency can vary this SSFS over several THz \cite{Kippenebrg_PRL2016raman}, making the influence of center frequency shift on $f_r$ more important. Here, we measure the dependence of $f_r$ on the SSFS via tuning of the pump frequency within the single CS regime.

We study a SiN microresonator, with a 100 $\mu$m ring radius, 800$\times$2000 nm waveguide geometry and a loaded Q-factor of 2.4$\times$10$^6$, with which we are able to generate a Kerr comb comprising single cavity solitons at 227 GHz repetition rate. The device has a drop-port (with power coupling coefficient of 1.6$\times$10$^{-4}$). More details about the microresonator and the characterization of the CS can be found in \cite{Weiner_OE2016intracavity}. The drop-port allows convenient measurement of the intracavity power, defined as the total power including the pump line, without the strong, directly transmitted pump that is present at the through-port. Frequency-comb assisted spectroscopy is used to measure the line spacing of the generated comb \cite{Kippenberg_NP2009frequency,Weiner_OE2016thermal}. A cw diode laser is used to sweep across the Kerr comb, generating beat notes that are recorded by an oscilloscope after bandpass filtering. A femtosecond frequency comb is used for calibration of the laser frequency sweep. This method has been shown to have MHz resolution and accuracy similar to that achieved using with the electro-optic modulation–assisted measurement of the line spacing  \cite{Kippenberg_NP2009frequency,Weiner_OE2016thermal}.


\begin{figure}[t]
\centering
\includegraphics[width=\linewidth]{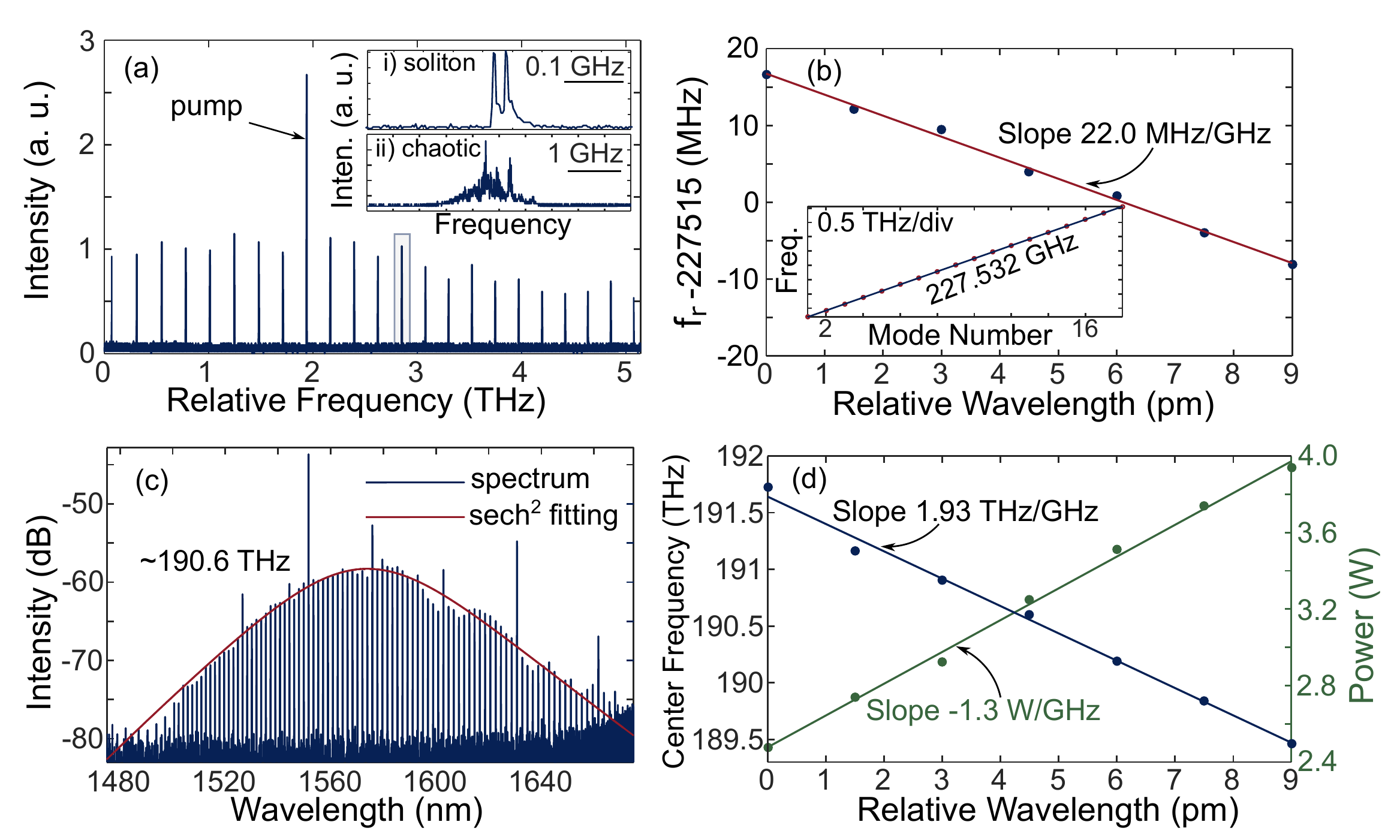}
\caption{(a) The frequency markers from the beat between the Kerr comb and the sweeping cw laser. Inset \lowercase\expandafter{\romannumeral1}) shows the zoom in a single frequency marker in the shaded box in the soliton state, while inset \lowercase\expandafter{\romannumeral2}) shows the frequency marker in the chaotic state. The frequency axis is calibrated by the frequency markers beating with the femtosecond frequency comb. (b) The dependence of $f_r$ on pump wavelength and the inset shows a typical linear fitting of the Kerr comb frequency markers to get $f_r$. (c) The soliton spectrum from the drop port and its sech$^2$ fit. (d) The measured dependence of the center frequency shift (blue dots) and the intracavity power (green dots) on the pump wavelength and their corresponding linear fits.}
\label{Fig2RepRate}
\end{figure}

As an example, the relative frequencies for different comb lines are shown in Fig. \ref{Fig2RepRate}(a). When we zoom into a single line in the CS regime, we can see two narrow peaks, since the cw laser generates a beat note that can pass the bandpass filter both when it is to the red and to the blue of comb line (inset (\lowercase\expandafter{\romannumeral1}) in Fig. \ref{Fig2RepRate}). In contrast, if the Kerr comb is in the chaotic regime, the beat with the cw laser is broad and structured (inset (\lowercase\expandafter{\romannumeral2}) in Fig. \ref{Fig2RepRate}(a)). We can fit the measured relative frequencies of the Kerr comb lines to a line to get $f_r$ (inset of Fig. \ref{Fig2RepRate}(b)). The power and SSFS of the CS will also change when we tune the pump frequency \cite{Kippenberg_NP2014temporal,Kippenebrg_PRL2016raman}. For our device, the single CS can be maintained over a 1.1 GHz range of pump frequency; $f_r$ changes by $\sim$25 MHz in this range, with a slope of 22.0 MHz/GHz (Fig. 1(b)). Note that this slope for a 18 GHz SiN microresonator based Kerr comb was measured to be 57 kHz/GHz for a non-soliton state in \cite{Wong_SR2015low}.

We also measure the spectrum of the generated CS, which fits well to a sech$^2$ function, except some distortion from the mode-interaction (see Fig. \ref{Fig2RepRate}(c)). The fit gives the center frequency of the CS. When the pump frequency varies, the center frequency of the CS also changes nearly linearly with the pump frequency \cite{Kippenebrg_PRL2016raman} (Fig. \ref{Fig2RepRate}(d)), having a slope of 1.93 THz/GHz. Using the drop port, we are also able to measure the average intracavity power; the power decreases nearly linearly with pump frequency, with a slope of $-$1.3 W/GHz.



\begin{figure}[t]
\centering
\includegraphics[width=0.75\linewidth]{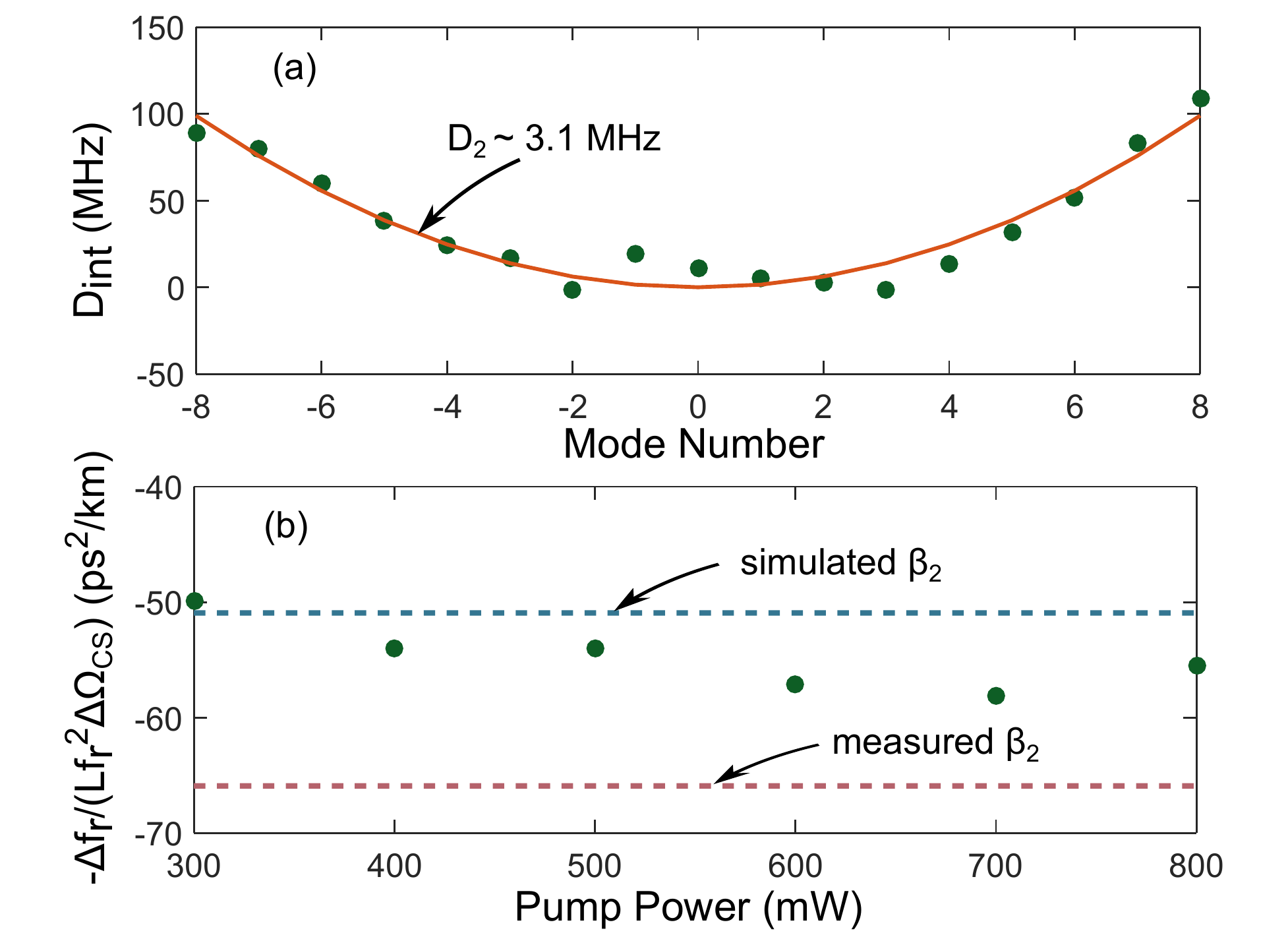}
\caption{(a) Resonance deviation from even spacing (circles) measured using frequency comb assisted spectroscopy and quadratic fit (line), yielding a dispersion of $-$66 ps$^2$/km. (b) The dispersion (circles) extracted from Eq. (1), based on the measurements of the SSFS and $\Delta f_r$ at different pump powers, assuming that the observed changes in repetition rate arisefrom SSFS only. These values are in close agreement with the dispersion measured using comb-assisted spectroscopy and computed via waveguide simulations.}
\label{Fig4Dis}
\end{figure}

The measurement of SSFS and intracavity power allows comparison between the influence of SSFS and thermal effects on $\Delta f_r$. The SSFS contribution to repetition rate change, denoted $\Delta f_r^{s}$, can be calculated as

\begin{equation}
\Delta {{f}_{r}^{s}}=\Delta \left( \frac{1}{{{\beta }_{1}}L} \right)=-\frac{L\Delta {{\beta }_{1}}}{{{\left( {{\beta }_{1}}L \right)}^{2}}}=-Lf_{r}^{2}{{\beta }_{2}} \Delta \Omega_{cs},
\label{EqDis}
\end{equation}
where $L$ is the length of the cavity, $1/\beta_1$ is the group velocity, $\beta_{2}$ is the group velocity dispersion, $\Omega_{cs}$ is the SSFS in angular frequency. The spectral recoil from mode-interaction in a MgF$_2$ cavity affects $f_r$ in a similar way \cite{Kippenberg_arXiv2016study}. From Eq. \ref{EqDis}, we can see $\Delta f_r^s$ depends on dispersion of the cavity for a given SSFS. To check this relation, we measure the dispersion using frequency comb assisted spectroscopy \cite{Kippenberg_NP2009frequency}. We show the deviation of the resonance from the equidistant spacing, $D_{int}=\nu_\mu-\nu_0-\mu D_1\approx D_2 \mu^2/2$, ($\nu_\mu$ is the resonant frequency of mode $\mu$, $D_1$ is the free spectral range, $D_2$ is the dispersion coefficient) in Fig. \ref{Fig4Dis}(a). From a quadratic fit, $D_2$ is found to be 3.1 MHz, equivalent to $\beta_2=-$66 ps$^2$/km \cite{Vahala_OL2016theory}, close to the simulated dispersion of $-$51 ps$^2$/km \cite{Weiner_OE2016intracavity}.

If we assume $\Delta f_r$ in Fig. 1(b) is driven solely by SSFS, we can extract $\beta_2$ by using Eq. 1. For an on-chip pump power of 800 mW, we obtain $\beta_2$ is -55 ps$^2$/km. Furthermore, the obtained value remains nearly the same for CSs generated under different pump powers (see Fig. \ref{Fig4Dis}(b)). More importantly, these values are quite close to those obtained via waveguide simulations and via linear spectroscopy measurements.  This agreement suggests that the chance in repetition rate is dominated by SSFS.

Furthermore, SSFS (denoted $\Omega_{CS}$) can be approximated as \cite{Vahala_OL2016theory},

\begin{equation}
{{\Omega }_{cs}}=\frac{8c{{\tau }_{R}}Q}{15{{n}_{g}}{{\omega }_{r}}}\frac{{{\beta }_{2}}}{\tau _{s}^{4}}=\frac{8{{\tau }_{R}Q}}{15\omega_r}\frac{{{\beta }_{2}}}{{{\beta }_{1}}\tau _{s}^{4}},
\label{EqSSFS}
\end{equation}
where $\omega_r$, $\omega_{p}$ are the resonance frequency of the cavity and frequency of the pump laser, respectively; $\tau_R$ is the Raman time constant; $n_g$, $c$, and $Q$ are the effective group index, speed of light and quality factor respectively; $\tau_s$ is the pulse-width of the CS. The peak power (P$_0$) of CSs follows the relationship, $\gamma {{P}_{0}}={\left| {{\beta }_{2}} \right|}/{\tau _{s}^{2}}$, ($\gamma$ is the nonlinear coefficient); P$_0$ also scales linearly with pump frequency detuning \cite{Bao_PRA2015carrier},

\begin{small}
\begin{equation}
\gamma {{P}_{0}}\approx\frac{2\left( {{\omega }_{r}}-{{\omega }_{p}} \right){{t}_{R}}}{L}\approx2\left( {{\omega }_{r}}-{{\omega }_{p}} \right){{\beta }_{1}},
\label{EqPeak}
\end{equation}
\end{small}
where $t_R$ is the round-trip time. Combining Eqs. \ref{EqDis}, \ref{EqSSFS}, \ref{EqPeak} and defining the pump frequency detuning as $\omega_d=\omega_r-\omega_p$, we can get,
\begin{equation}
\Delta f_{r}^{s}\approx-\frac{32{{\tau }_{R}}}{15}\frac{f_rQ}{\omega_r}\Delta {{\omega }_{d}}^{2}\approx -\frac{64{{\tau }_{R}}}{15}\frac{{{\omega }_{d}Q}}{ {{\omega }_{r}}}{{f}_{r}}\Delta {{\omega }_{d}}.
\label{EqSSFSfr}
\end{equation}
This relationship means $\Delta f_r^s$/$\Delta\omega_{d}$ is immune to the dispersion or nonlinear coefficient of the cavity.

\begin{figure}[t]
\centering
\includegraphics[width=0.85\linewidth]{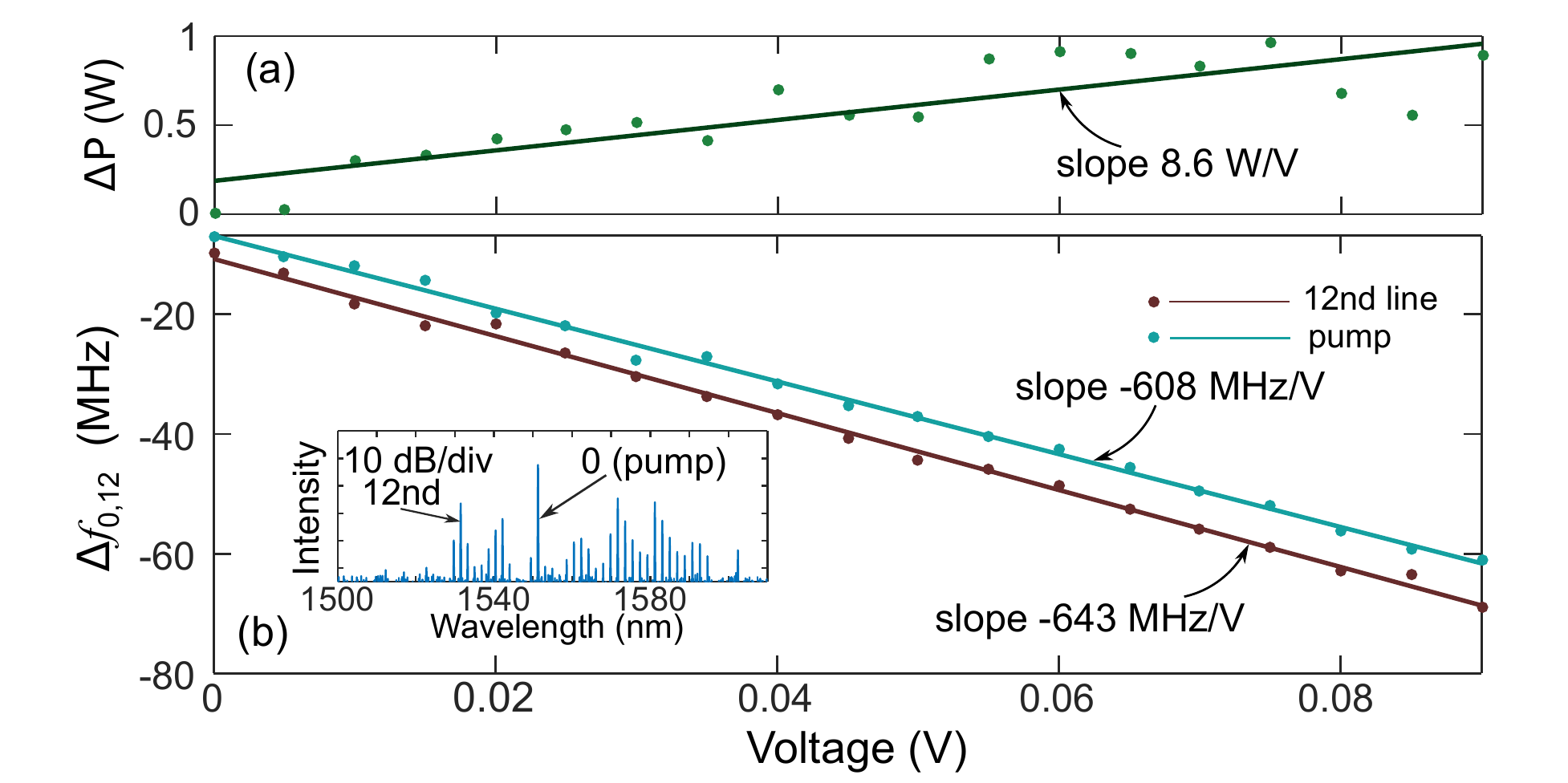}
\caption{(a) The measured intracavity power change and its linear fit versus the voltage to control the piezo of the pump laser. (b) The measured frequency change and the linear fit for the pump line and the 12nd comb line versus piezo voltage. The change of the frequency for 12nd comb line is shifted from zero for clarity.}
\label{Fig3Thermal}
\end{figure}

We now assess the contribution of the thermo-optic (TO) effect to the change in repetition rate, which we denote $\Delta f_r^t$. This contribution can be expressed as
\begin{equation}
\Delta f_{r}^{t}=\Delta \left( \frac{c}{n_gL} \right)=-\frac{L}{c}\left( \frac{{{c}^{2}}}{{{n_g}^{2}}{{L}^{2}}} \right)\Delta n_g=-\frac{Lf_{r}^{2}}{c}\Delta n_g,
\label{EqThermal}
\end{equation}
where $\Delta n_g$ is the change of the group index due to the TO effect. $\Delta n_g$ can be written,

\begin{equation}
\Delta {{n}_{g}}=\frac{d{{n}_{g}}}{dT}\frac{{{\tau }_{\theta }}{({\alpha }f_{\theta })}\Delta P}{{{C}_{\theta }}\rho {{A}_{eff}}}.
\label{EqIndex}
\end{equation}
Here d$n_g$/dT is the derivative of effective group index with respect to temperature. To get a value for d$n_g$/dT, we perform waveguide simulations to obtain the effective phase index ($n_p$) of the microresonator for different wavelengths and temperatures; the group index is $n_g=n_p- \lambda dn_p/d\lambda$ \cite{Weiner_OE2016thermal}. Using the same TO coefficient for SiN and SiO$_2$ as \cite{Weiner_OE2016thermal}, we find $dn_g/dT$=3.2$\times$10$^{-5}$ K$^{-1}$. $C_\theta$=760 J/(kg$\cdot$K) is the heat capacity of SiN, $\rho$=2.2$\times$10$^3$ kg/m$^3$ is the density of SiN \cite{Yeshaiahu_OE2008thermal}; the time constant ($\tau_\theta$) was measured to be 0.25 $\mu$s in a similar microresonator in our group's previous work \cite{Qi_CLEO2014fast}; $\alpha$ is the propagation loss coefficient (measured to be $1.7\times10^{-3}/628~\mu m^{-1}$), $f_\theta$ is the fraction of the absorbed energy that is converted to heat, $A_{eff}$ is the effective mode area (extracted to be 0.67 $\mu$m$^2$ from the simulation based on the geometry of the ring), $\Delta P$ is the change of average intracavity power. Note that we neglect thermal expansion in the calculation; however, the SiN ring is embedded in a 4 $\mu$m thick SiO$_2$ layer and the expansion should be small \cite{Weiner_OE2016thermal}. To get the upper bound estimate of $\Delta f_r^t$, we set $f_\theta=1$. Based on Eqs. \ref{EqThermal}, \ref{EqIndex} the thermal induced $\Delta f_r^t$ is 3.1 MHz for a measured intracavity power change of 1.5 W. Even though $f_\theta=1$ overestimates the contribution from the TO effect, the estimated value for $\Delta f_r^t$ is still an order of magnitude smaller than the observed 25 MHz change in comb spacing, suggesting that the contribution from the TO effect is much weaker than that of SSFS.

To further isolate the thermal contribution, we conduct another measurement of $\Delta f_r$ in the stable modulation instability regime (see the inset of Fig. 3(b) for a typical spectrum), where the Kerr comb does not experience SSFS. The pump line and the 12th  line are filtered by a pulse-shaper to beat with the femtosecond comb, thus measuring 12$\Delta f_r$. We record both the change of the frequencies of these two Kerr comb lines and the average comb power with respect to the voltage used to control the piezo of the external cavity diode pump laser in Fig. \ref{Fig3Thermal}. Both the frequencies and the comb power are found to change nearly linearly. From the linear fits we find that the coefficient of repetition rate change with intracavity power is $-$0.3 MHz/W. Hence, for the measured 1.5 W change in intracavity power in Fig. \ref{Fig2RepRate}(d), the contribution to $\Delta f_r$ is only 0.45 MHz, much smaller than the change in repetition rate observed.


Since $\Delta f_r$ is dominated by SSFS, which can be modeled by the generalized LLE \cite{Kippenebrg_PRL2016raman,Vahala_Optica2015soliton,Vahala_OL2016theory}, we perform simulations using the generalized LLE to look into the pulse timing dynamics. We write the generalized LLE \cite{Coen_OL2013modeling,Kippenebrg_PRL2016raman,Bao_OE2015soliton} as:

\begin{small}
\begin{equation}
\begin{aligned}
  & \left( {{t }_{R}}\frac{\partial }{\partial t}+\frac{\kappa_0 +\kappa_1 }{2}+i{{\delta }_{0}}+i\frac{{{\beta }_{2}}L}{2}\frac{{{\partial }^{2}}}{\partial {{\tau }^{2}}} \right)E-\sqrt{\kappa_1 }{{E}_{in}} \\
 & -i\gamma L \left( 1+\frac{i}{{{\omega }_{0}}}\frac{\partial }{\partial \tau } \right)\left( E\int_{-\infty }^{+\infty }{R({\tau }')\left| E(t,\tau -{\tau }') \right|^2 d{\tau }'} \right)=0, \\
\end{aligned}
\label{EqLLE}
\end{equation}
\end{small}
where $t_R$ is the round-trip time, $R(\tau )=(1-{\Theta_{R}})\delta (\tau )+{\Theta_{R}}{{h}_{R}}(\tau )$ is the nonlinear response, including both the electronic and the delayed Raman response ($h_R(\tau)$), $\Theta_R$ is the Raman fraction (chosen as 0.13), $\kappa_0$ and $\kappa_1$ are the intrinsic loss and the external coupling coefficient respectively and $E_{in}$ is the pump field at the frequency $\omega_0$.

\begin{figure}[t]
\centering
\includegraphics[width=0.9\linewidth]{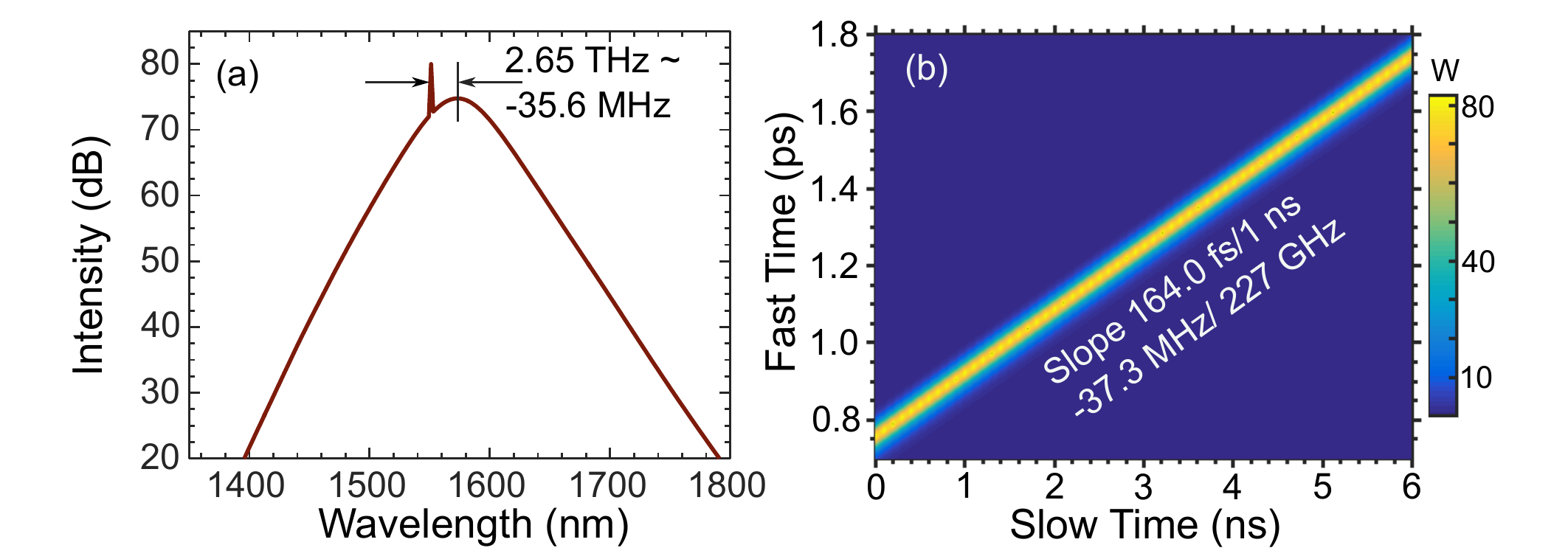}
\caption{(a) The simulated spectrum of the intracavity CS, showing a SSFS of $-2\pi\times$2.65 THz. (b) The temporal evolution of the CS train on slow time. The peak of the temporal evolution trace has a slope of 164 fs/1 ns.}
\label{Fig5Sim}
\end{figure}

Here, we show simulation results for a typical CS generated at 305 mW pump power and detuning $\delta_0$=0.0303 in a SiN microresonator with $\gamma$=0.9 (W$\cdot$ m)$^{-1}$, $\beta_2=-66$ ps$^2$/km, $\kappa_0$=0.002 and $\kappa_1$=0.0004. The spectrum of the CS is shown in Fig. \ref{Fig5Sim}(a), corresponding to a pulse-width $\tau_s$=50 fs. The shift between the center frequency of the spectrum and the pump frequency, or SSFS, is $-2\pi\times$2.65 THz. According to Eq. \ref{EqDis}, this leads to a change of repetition rate of $-$35.6 MHz for $f_r$=227 GHz. To compare this value with the pulse timing dynamics in the LLE simulation, we show the temporal-spatial dynamics of the simulated CS in Fig. \ref{Fig5Sim}(b). Due to SSFS, the peak of the CS envelope experiences additional delay with a slope of 164.0 fs/ns; without SSFS the plot would appear as a horizontal line. From the relationship  $-\Delta t_R/t_R = \Delta f_r/ f_r$, this slope is equivalent to $\Delta f_r$ =$-$37.2 MHz, in close agreement with the SSFS predicted $-$35.6 MHz. This supports the notion that pulse timing information is retained in the LLE, despite the inherent averaging in the derivation.

The slight difference in the two values of $\Delta f_r$ can be attributed to the inclusion of Kerr shock (KS) in the simulation, which affects the group velocity of the CS \cite{Haus_OL2001group,Cundiff_OL2007quantitative,Cundiff_OL2014pulse,Bao_PRA2015carrier}. The difference in the simulated and the SSFS predicted $\Delta f_r$ is equivalent to a change of round-trip time of $\Delta t_R/t_R$=7.2 fs/ns. Since the change of group velocity ($v_g$) for sech-soliton induced by KS is $\Delta(1/v_g)\approx\gamma P_{0}/\omega_0$ \cite{Haus_OL2001group}, the variation of the round-trip time induced by KS yields,

\begin{equation}
  \frac{\Delta {{t }_{R}}}{{{t }_{R}}}=\frac{\Delta ({L}/{{{v}_{g}}}\;)}{{{t }_{R}}}=\Delta \left( \frac{1}{{{v}_{g}}} \right)\frac{L}{{{t }_{R}}}\approx\frac{\gamma P_{0}L}{{{\omega }_{0}}{{t }_{R}}}.
\label{EqKS}
\end{equation}
Based on Eq. \ref{EqKS}, $\Delta t_R/t_R$ induced by KS is 8.8 fs/ns, which is close to the simulated 7.2 fs/ns. The residual discrepancy may result from the deviation of the simulated CS from the sech-pulse. Furthermore, when excluding KS in the simulation, the SSFS increases slightly to $-2\pi\times2.73$ THz. The round-trip time change predicted from this SSFS is $\Delta t_R/t_R$=161.6 fs/ns based on Eq. 1. On the other hand, if we extract the timing of the CS envelope peak from the simulation without KS, we find a slope of 161.7 fs/ns, in excellent agreement with the SSFS prediction. Hence, the generalized LLE can be used to model pulse timing and changes in repetition rate induced both by SSFS and by KS.

Since the pulse dynamics in nonlinear resonators strongly affect the comb performance \cite{Cundiff_OL2007quantitative,Cundiff_OL2014pulse,Cundiff_PRL2015observation}, an accurate theoretical model to investigate the intracavity pulse dynamics is important to the optimization of frequency combs. For mode-locked lasers, the complex Ginzburg-Landau equation is usually used to model the pulse dynamics \cite{Haus_JQE1975theory}. However, effects such as gain relaxation, spectral filtering and the resulting mode-locking state etc. are usually hard to model accurately. Hence, investigation of pulse dynamics in mode-locked laser usually centers on experiment. For CSs in passive microresonators, the absence of active gain and nonlinear absorption makes it feasible to model the pulse timing dynamics accurately. If noise sources are added to the model correctly, we can expect a full characterization of the pulse dynamics.

In summary, the SSFS induces a considerable change in the repetition rate of Kerr combs operating in the single cavity soliton regime. Both experimental and modeling results suggest that the SSFS can lead to changes in repetition rate much larger than caused by the thermo-optic effect.  Although the SSFS does make the repetition rate of CS combs more sensitive to changes in pump frequency than other, non-soliton coherent Kerr comb states, it may also provide a useful degree of freedom to broaden the locking range over which the repetition rate may be stabilized.  Ultimately, understanding the pulse dynamics and noise behavior of soliton Kerr combs based on the LLE can be valuable for the improvement of Kerr comb stability.

\section*{Acknowledgment}
Air Force Office of Scientific Research (AFOSR) (FA9550-15-1-0211), DARPA PULSE program (W31P40-13-1-0018) from AMRDEC, and National Science Foundation (NSF) (ECCS-1509578).

\bibliography{reflist2}
\parskip 12pt
\end{document}